\documentclass[preprint,showpacs,amsmath,nofootinbib]{revtex4}

\begin{document}

\preprint{}

\title{Dynamics of Gravity as Thermodynamics on the Spherical Holographic Screen}

\author{Yu Tian}
\affiliation{College of Physical Sciences, Graduate University of
Chinese Academy of Sciences, Beijing 100049, China}
\affiliation{Kavli Institute for Theoretical Physics China, CAS,
Beijing 100190, China}

\author{Xiao-Ning Wu}
\affiliation{Institute of Mathematics, Academy of Mathematics and
System Science, {CAS}, Beijing 100190, China} \affiliation{Hua
Loo-Keng Key Laboratory of Mathematics, {CAS}, Beijing 100190,
China}

\date{\today}

\begin{abstract}
The dynamics of general Lovelock gravity, viewed on an arbitrary
spherically symmetric surface as a holographic screen, is recast as
the form of some generalized first law of thermodynamics on the
screen. From this observation together with other two distinct
aspects, where exactly the same temperature and entropy on the
screen arise, it is argued that the thermodynamic interpretation of
gravity is physically meaningful not only on the horizon, but also
on a general spherically symmetric screen.
\end{abstract}

\pacs{04.70.Dy, 04.50.-h, 04.20.Cv}

\maketitle


The discovery of black hole entropy and thermodynamics \cite{thermo}
reveals a rather general and profound relation between gravity and
thermodynamics. Later, based on the area law of entropy for all
local acceleration horizons, Jacobson derived the Einstein equations
from the first law of thermodynamics \cite{Jacobson}. This
derivation can be extended to non-Einstein gravity (for a review,
see \cite{review}). Recently, Padmanabhan reinterpreted the relation
$E=2T S$ \cite{Padmanabhan1} between the Komar energy, temperature
and entropy as the equipartition rule of energy \cite{Padmanabhan2},
and Verlinde proposed the gravity as an entropic force and derived
the Einstein equations \cite{Verlinde} from the equipartition rule
of energy and the holographic principle
\cite{holographic,holography}. For an incomplete list of related
works, see \cite{following,another,other,Tian,Chen,Cai}.

Perhaps one of the most important things that we learn from these
recent works is that there should be certain thermodynamic
interpretation of the gravitational dynamics on some general surface
as a holographic screen, instead of the well-known thermodynamic
relations or interpretations found only on various kinds of
horizons. In fact, on a general screen in a static space-time, there
is a locally defined Unruh-Verlinde temperature \cite{Verlinde},
which has clear physical meaning and appears in the equipartition
rule of energy. On the other hand, Padmanabhan has generalized the
definition of entropy on the horizon by Wald et al \cite{Wald} to
the off-horizon case in a certain class of gravitational theories,
which always gives one quarter of the screen's area in Einstein's
gravity \cite{another}. Many other authors also suggest $S=A/4$ for
some general holographic screen from different aspects \cite{other}.
{(See, however, \cite{Tian} for a different entropy formula.)}
Furthermore, recall that there are various definitions of
quasi-local mass (energy) associated to the region enclosed by the
screen, besides the Tolman-Komar energy appearing in the
equipartition rule. So, it is time to put together all the pieces of
thinking and see whether they can fit into a whole picture. Chen et
al have made an attempt to this direction in four dimensional
Einstein's gravity and obtained a generalized first law of
thermodynamics for the spherically symmetric screen \cite{Chen}, but
the energy appearing there is the Arnowitt-Deser-Misner (ADM) mass
instead of some quasi-local energy associated to the screen. Even
earlier, Cai et al have also considered the spherically symmetric
case in Einstein's gravity and obtained a relation similar to the
generalized first law \cite{Cai} (see e.g. \cite{on-horizon} for the
on-horizon case), but that is a dynamical process, while in the
usual thermodynamic sense the generalized first law should describe
the quasi-static processes. Although most of the present works
support some thermodynamic relations on a general screen with
$S=A/4$ in Einstein's gravity, this entropy formula seems too simple
to be falsified. Therefore, it is necessary to investigate more
general theories of gravity, where expressions of the entropy and
other quantities are complicated enough, and to collect more,
different evidence for supporting that conclusion.

In this letter, we consider a general spherically symmetric screen
in the Lovelock gravity in arbitrary dimensions \cite{Lovelock}, one
of the most natural extensions of Einstein's gravity. Some
thermodynamic interpretations or relations\footnote{See
\cite{non-Einstein} for some related works on the horizon
thermodynamics in non-Einstein gravity.} on the screen are
investigated from three distinct aspects:
\begin{itemize}
  \item The gravitational equations of motion are reinterpreted as a
  generalized first law, which involves some Misner-Sharp-like energy inside the screen.
  \item The analysis in \cite{Chen} is generalized to this case,
  which for {the Reissner-Nordstr\"{o}m solution in Einstein's gravity} involves the Tolman-Komar energy inside the screen after a
  Legendre transformation.
  \item Padmanabhan's general definition of entropy on the screen is
  explicitly computed, which satisfies some equipartition-like rule.
\end{itemize}
In all these aspects, exactly the same entropy and Unruh-Verlinde
temperature arise, so it is convincing that the quantities and
thermodynamic interpretations on the screen are physically
meaningful.

\vspace{0.5cm}

Take Einstein's gravity in $n$ space-time dimensions as the simplest
example to illustrate our basic strategy. The action functional is
\begin{equation}
S=\int d^n x (\frac{\sqrt{-g}}{16\pi} R+\mathcal{L}_\mathrm{matt}),
\end{equation}
which leads to the equations of motion
\begin{equation}\label{EOM}
R_{ab}-\frac 1 2 R g_{ab}=8\pi T_{ab}.
\end{equation}
The most general static, spherically symmetric metric can be written
as
\begin{equation}\label{general_static}
ds^2=-e^{-2c(r)} f(r) dt^2+f(r)^{-1} dr^2+r^2 d\Omega_{n-2}^2
\end{equation}
with $d\Omega_{n-2}^2$ the metric on the unit $(n-2)$-sphere. {At
present}, we assume the ansatz
\begin{equation}\label{ansatz}
ds^2=-f(r) dt^2+f(r)^{-1} dr^2+r^2 d\Omega_{n-2}^2
\end{equation}
for the metric, which means that the Lagrangian density
$\mathcal{L}_\mathrm{matt}$ of matter cannot be too arbitrary, while
still containing many cases of physical interest, such as
electromagnetic fields, the cosmological constant and {some kinds of
quintessential matter} \cite{fluids}, etc. {The general case
(\ref{general_static}) will be discussed finally. In the space-time
(\ref{ansatz})}, the Unruh-Verlinde temperature on the spherical
screen of radius $r$ is easily obtained as
\begin{equation}\label{Unruh}
T=\frac{-\partial_r g_{tt}}{4\pi\sqrt{-g_{tt}
g_{rr}}}=\frac{f'}{4\pi},
\end{equation}
which is purely geometric and so independent of the gravitational
dynamics. Here a prime means differentiation with respect to $r$.

Upon substitution of the ansatz (\ref{ansatz}) into the equations of
motion (\ref{EOM}), the nontrivial part of them is
\cite{Padmanabhan}
\begin{equation}\label{non-trivial}
r f'-(n-3)(1-f)=\frac{16\pi P}{n-2} r^2
\end{equation}
with $P=T^r_r=T^t_t$ the radial pressure. Now we focus on a
spherical screen with fixed $f$ \cite{Chen} in different static,
spherically symmetric solutions of (\ref{EOM}). In order to do so,
we just need to compare two such configurations of infinitesimal
difference.{\footnote{This is the standard point of view in
thermodynamics, as well as in black-hole thermodynamics, which
considers the variation among continuously many static
configurations, focusing on the horizon (as a surface with fixed
$f=0$ but varied $r$) of each configuration.}} In fact, multiplying
both sides of (\ref{non-trivial}) by the factor
\begin{equation}\label{factor}
\frac{n-2}{16\pi}\Omega_{n-2} r^{n-4} dr,
\end{equation}
we have after some simple algebra (assuming $f$ fixed)
\begin{eqnarray}
&&\frac{f'}{4\pi} d\left(\frac{\Omega_{n-2}
r^{n-2}}{4}\right)-d\left(\frac{n-2}{16\pi}\Omega_{n-2} (1-f)
r^{n-3}\right)\nonumber\\
&=&P d(\frac{\Omega_{n-2} r^{n-1}}{n-1}).
\end{eqnarray}
The above equation is immediately recognized as the generalized
first law
\begin{equation}\label{first_law}
T dS-dE=PdV
\end{equation}
with $T$ the Unruh-Verlinde temperature (\ref{Unruh}) on the screen,
\begin{eqnarray}
S&=&\frac{\Omega_{n-2} r^{n-2}}{4},\label{entropy}\\
E&=&\frac{n-2}{16\pi}\Omega_{n-2} (1-f) r^{n-3}\label{energy}
\end{eqnarray}
and $V=\frac{\Omega_{n-2} r^{n-1}}{n-1}$ the volume of the
(standard) unit $(n-1)$-ball. Here $E$ is just the standard form of
the Misner-Sharp energy inside the screen in spherically symmetric
space-times \cite{M-S}, which is also identical to the
Hawking-Israel energy in this case. More explicitly, solving $f$
from (\ref{energy}) gives
\begin{equation}
f=1-\frac{16\pi E}{(n-2)\Omega_{n-2} r^{n-3}},
\end{equation}
which is the Schwarzschild solution in $n$ dimensions for constant
$E$ as its ADM mass, and some general spherically symmetric solution
for certain mass function $E(r)$.

Some remarks are in order. First, the generalized first law
(\ref{first_law}) is of the same form as that in \cite{Padmanabhan}
for the horizon of spherically symmetric black holes, but is valid
for general spherical screen with fixed $f$, which includes the
horizon as the special case $f=0$. Second, the entropy
(\ref{entropy}) in the generalized first law is actually $S=A/4$,
i.e. one quarter of the area, for a general spherical screen, the
same as the result obtained in \cite{Chen} by the generalized
Smarr's approach for the four dimensional case. (Similar results
appear in \cite{Cai} and \cite{another}, as mentioned previously.)
In fact, the generalized Smarr's approach can be used in the higher
dimensional case without any difficulty. The Reissner-Nordstr\"{o}m
solution in $n$ dimensions is
\begin{equation}
f=1-\frac{2\mu}{r^{n-3}}+\frac{q^2}{r^{2n-6}},
\end{equation}
where the mass parameter $\mu$ is related to the ADM mass $M$ by
$\mu=\frac{8\pi M}{(n-2)\Omega_{n-2}}$. Upon generalization of
Smarr's approach (fixing $f$), the above expression leads to
\cite{preparation}
\begin{equation}\label{ADM}
dM=T dS+\phi dq
\end{equation}
with $T$ the Unruh-Verlinde temperature (\ref{Unruh}) on the screen,
$S$ again given by (\ref{entropy}) and
\begin{equation}\label{phi}
\phi=\frac{(n-2)\Omega_{n-2} q}{8\pi r^{n-3}}
\end{equation}
proportional to the electrostatic potential \emph{on the screen}.
Furthermore, by straightforwardly working out the Tolman-Komar
energy $K=M-\phi q$ inside the screen, which is just a Legendre
transformation of $M$, we can obtain another generalized first law
\begin{equation}\label{Komar}
dK=T dS-q d\phi,
\end{equation}
which seems even more relevant to the holographic picture, since now
all the quantities are closely related to the screen, and the
Tolman-Komar energy $K$ satisfies the equipartition rule
\cite{Padmanabhan2,Verlinde}. {Anyway,} exactly the same temperature
$T$ and entropy $S$ appear in different kinds of generalized first
laws\footnote{There is no obvious relation between (\ref{ADM}) [or
(\ref{Komar})] with (\ref{first_law}), as can be seen more clearly
in the discussion around (\ref{relation}) for the general Lovelock
gravity.} and other places such as \cite{Cai} and \cite{another},
which is strong evidence that the Unruh-Verlinde temperature
(\ref{Unruh}) and the entropy (\ref{entropy}) should make sense in
physics. This argument will be further confirmed in more general
cases below.

\vspace{0.5cm}

Now we consider the general Lovelock gravity. The action functional
is
\begin{equation}\label{Lovelock_action}
S=\int d^n x (\frac{\sqrt{-g}}{16\pi}\sum_{k=0}^m\alpha_k
L_k+\mathcal{L}_\mathrm{matt})
\end{equation}
with $\alpha_k$ the coupling constants and
\begin{equation}
L_k=2^{-k}\delta^{a_1 b_1\cdots a_k b_k}_{c_1 d_1\cdots c_k d_k}
R_{a_1 b_1}^{c_1 d_1}\cdots R_{a_k b_k}^{c_k d_k},
\end{equation}
where $\delta^{ab\cdots cd}_{ef\cdots gh}$ is the generalized delta
symbol which is totally antisymmetric in both sets of indices. Note
that $\alpha_0$ is proportional to the cosmological constant and
$L_1=R$. By the ansatz (\ref{ansatz}) again and extending the
approach in \cite{Wheeler} for the vacuum case to include
$\mathcal{L}_\mathrm{matt}$, the nontrivial part of the equations of
motion is
\begin{equation}\label{Gtt}
\sum_k\tilde{\alpha}_k (\frac{1-f}{r^2})^{k-1}[ k r
f'-(n-2k-1)(1-f)]=\frac{16\pi P}{n-2} r^2,
\end{equation}
where
\begin{equation}
\tilde{\alpha}_0=\frac{\alpha_0}{(n-1)(n-2)},\quad\tilde{\alpha}_1=\alpha_1,\quad\tilde{\alpha}_{k>1}=\alpha_k\prod_{j=3}^{2k}(n-j).
\end{equation}
Now follow the strategy illustrated in the Einstein case.
Multiplying both sides of the above equation by the factor
{(\ref{factor}),} we have after some simple algebra (assuming $f$
fixed)
\begin{eqnarray}
&&\frac{f'}{4\pi} d\left(\frac{n-2}{4}\Omega_{n-2} r^{n-2} \sum_k
\frac{\tilde{\alpha}_k k}{n-2k} (\frac{1-f}{r^2})^{k-1}\right)\nonumber\\
&&-d\left(\frac{n-2}{16\pi}\Omega_{n-2}
r^{n-1}\sum_k\tilde{\alpha}_k
(\frac{1-f}{r^2})^k\right)\nonumber\\
&=&P d(\frac{\Omega_{n-2} r^{n-1}}{n-1}).\label{explicit_1st_law}
\end{eqnarray}
Recalling that the Unruh-Verlinde temperature (\ref{Unruh}) is
independent of {dynamics}, we again recognize the above equation as
the generalized first law (\ref{first_law}) with
\begin{eqnarray}
S&=&\frac{n-2}{4}\Omega_{n-2} r^{n-2} \sum_k \frac{\tilde{\alpha}_k
k}{n-2k} (\frac{1-f}{r^2})^{k-1},\label{Lovelock_entropy}\\
E&=&\frac{n-2}{16\pi}\Omega_{n-2} r^{n-1}\sum_k\tilde{\alpha}_k
(\frac{1-f}{r^2})^k.\label{Lovelock_energy}
\end{eqnarray}
Here $E$ can be interpreted as some generalization of the
Misner-Sharp (or Hawking-Israel) energy to the Lovelock gravity (for
certain special case, see \cite{Lovelock_M-S} for the discussion of
the Misner-Sharp energy). In fact, when $E=M$ is a constant,
(\ref{Lovelock_energy}) is just the algebraic equation (of arbitrary
degree) that $f$ satisfies for the vacuum case \cite{Wheeler}, with
$M$ the ADM mass. And when
\begin{equation}\label{RN-like}
E(r)=M-\frac{\phi(r) q}{2}
\end{equation}
with $\phi(r)$ given by (\ref{phi}), (\ref{Lovelock_energy}) gives
the Reissner-Nordstr\"{o}m-like solution with charge $q$, while the
Born-Infeld-like case corresponds to more complicated mass function
$E(r)$ \cite{Lovelock-RN}.

The entropy (\ref{Lovelock_entropy}) should be discussed further. On
the horizon, we have $f=0$, so (\ref{Lovelock_entropy}) just becomes
the well-known entropy of {Lovelock black holes} \cite{Lovelock_BH}.
On a general spherical screen with fixed $f$, the generalized
Smarr's approach can be applied without knowing the explicit form of
$f$ (which is impossible in the general Lovelock gravity) and still
gives {the generalized first law (\ref{ADM})} for the
Reissner-Nordstr\"{o}m-like solution (\ref{RN-like}), with exactly
the same temperature (\ref{Unruh}) and entropy
(\ref{Lovelock_entropy}) \cite{preparation}. Here the key point is
that the entropy (\ref{Lovelock_entropy}) and energy
(\ref{Lovelock_energy}) satisfy
\begin{equation}\label{relation}
\frac{\partial S}{\partial r}=-4\pi\frac{\partial E}{\partial f},
\end{equation}
which is independent of the previous interpretation of
(\ref{explicit_1st_law}) as the generalized first law
(\ref{first_law}). Furthermore, Padmanabhan has proposed another
definition of entropy off the horizon \cite{another} in a certain
class of theories including the Lovelock gravity, generalizing the
definition of entropy on the horizon by Wald et al \cite{Wald}. For
a general screen $\mathcal{S}$, the associated entropy is suggested
to be
\begin{equation}\label{general}
S=\int_\mathcal{S} 8\pi
P^{ab}_{cd}\epsilon_{ab}\epsilon^{cd}\sqrt{\sigma} d^{n-2} x
\end{equation}
with
\begin{equation}\label{P}
P^{abcd}=\frac{\partial L}{\partial R_{abcd}},
\end{equation}
$\epsilon_{ab}$ the binormal to $\mathcal{S}$ and $\sigma_{ab}$ the
metric on $\mathcal{S}$, where $L=\frac{1}{16\pi}\sum_k\alpha_k L_k$
for the Lovelock gravity. This entropy is shown to satisfy the
equipartition-like rule with the Unruh-Verlinde temperature
(\ref{Unruh}) and some generalized Tolman-Komar energy
\cite{another}. In our case, the only non-vanishing components of
the binormal are $\epsilon_{tr}=1/2=-\epsilon_{rt}$, so the only
relevant component of (\ref{P}) in (\ref{general}) is
\begin{equation}\label{Ptrtr}
P^{tr}_{tr}\sim\sum_k\alpha_k k 2^{-k}\delta^{tr a_2 b_2\cdots a_k
b_k}_{tr c_2 d_2\cdots c_k d_k} R_{a_2 b_2}^{c_2 d_2}\cdots R_{a_k
b_k}^{c_k d_k}
\end{equation}
with the indices $a_i,b_i,c_i,d_i$ ($i=2,\cdots,k$) running only
among the angular directions. By explicitly working out
\begin{equation}\label{R}
R^{ab}_{cd}=\frac{1-f}{r^2}\delta^{ab}_{cd}
\end{equation}
for the metric (\ref{ansatz}) and substituting it into
(\ref{Ptrtr}), one can see that {(\ref{general})} eventually gives
(\ref{Lovelock_entropy}) \cite{preparation}.

\vspace{0.5cm}

{For the general case (\ref{general_static}) in the Lovelock
gravity, it turns out that the nontrivial part of the equations of
motion includes
\begin{equation}\label{minus}
\sum_k\tilde{\alpha}_k (\frac{1-f}{r^2})^{k-1} (2k r f
c')=\frac{16\pi (T_t^t-T_r^r)}{n-2} r^2
\end{equation}
in addition to (\ref{Gtt}) with $P$ replaced by $T_t^t$. Taking a
linear combination of these two equations and multiplying both sides
by the factor (\ref{factor}), we have again the generalized first
law (\ref{first_law}) with exactly the same entropy
(\ref{Lovelock_entropy}) and Misner-Sharp energy
(\ref{Lovelock_energy}), but with a slightly different temperature
\begin{equation}\label{gen_temperature}
T=\frac{f'-2l f c'}{4\pi}=\frac{\partial_r [(g^{rr})^{1-l}
(-g_{tt})^l]}{4\pi (-g_{tt} g_{rr})^l}
\end{equation}
and $ P=(1-l) T_t^t+l T_r^r$. In fact, the choice $l=1/2$ just gives
the generalization of Hayward's approach to the Lovelock gravity
\cite{preparation}. Another choice $l=1$ is of special interest,
since in this case $P=T_r^r$ is just the standard expression of the
radial pressure and
\begin{equation}\label{mod_temperature}
T=\frac{\partial_r g_{tt}}{4\pi g_{tt} g_{rr}}
\end{equation}
differs from the standard Unruh-Verlinde temperature only by a
$\sqrt{-g_{tt} g_{rr}}$ factor. Similar phenomena of non-unique
temperatures are extensively observed in the on-horizon case
\cite{non-unique,Cai}. Furthermore, the definition (\ref{general})
of entropy still gives (\ref{Lovelock_entropy}), since (\ref{R})
holds even in this case.}

Nevertheless, there are many open questions and/or unclear points in
this framework, of which an important one will be described as
follows, simply in Einstein's gravity. Although the relation $S=A/4$
for a general spherically symmetric screen seems rather
universal\footnote{The same relation even holds in the holographic
viewpoint of entanglement entropy \cite{entanglement}.} and is
supported by many recent works, there is an alternative expression
$S=2\pi R E$ obtained in \cite{Tian}, which seems also substantial.
In fact, the former form of entropy just saturates the holographic
entropy bound \cite{holographic}, while the latter form just
saturates the Bekenstein entropy bound \cite{Bekenstein}.
Furthermore, for the former form of entropy it is easy to write down
some generalized first law of thermodynamics as discussed above, but
it is not clear how to realize Verlinde's entropy variation formula
and then the gravity as an entropic force, while for the latter form
there exist the entropy variation formula and the entropic force
expression \cite{Tian} but without a satisfactory generalized first
law. How to reconcile these two forms of entropy is a significant
open question.

\begin{acknowledgments}
We thank Prof. R.-G. Cai, C.-G. Huang, Y. Ling, C.-J. Gao and Dr.
H.-B. Zhang for helpful discussions. This work is partly supported
by the National Natural Science Foundation of China (Grant Nos.
10705048, 10731080 and {11075206}) and the President Fund of GUCAS.
\end{acknowledgments}

\end{document}